\documentclass[twocolumn,aps,showpacs,preprintnumbers,nofootinbib,prd,superscriptaddress,groupedaddress,showkeys,10pt]{revtex4-1}
\usepackage{amsmath}
\usepackage{epsfig}
\usepackage{graphics}
\usepackage{color}
\usepackage{graphicx}
\usepackage[linktocpage]{hyperref}
\usepackage{enumerate}
\usepackage{tensor}
\usepackage{mathtools}
\usepackage{url}
\begin{document}

\title{Effect of the cosmological constant on the deflection angle by a rotating cosmic string}

\author{Kimet Jusufi$^{1,2}$}
\email{kimet.jusufi@unite.edu.mk}
\author{Ali \"{O}vg\"{u}n$^{3,4,5,6}$}
\email{ali.ovgun@pucv.cl}

\affiliation{${^1}$ Physics Department, State University of Tetovo, Ilinden Street nn, 1200,
Tetovo, Macedonia }
\affiliation{${^2}$ Institute of Physics, Faculty of Natural Sciences and Mathematics, Ss. Cyril and Methodius University, Arhimedova 3, 1000 Skopje, Macedonia}

\affiliation{${^3}$ Instituto de F\'{\i}sica, Pontificia Universidad Cat\'olica de
Valpara\'{\i}so, Casilla 4950, Valpara\'{\i}so, Chile }

\affiliation{${^5}$ Physics Department, Arts and Sciences Faculty, Eastern Mediterranean University, Famagusta, North Cyprus via Mersin 10, Turkey }

\affiliation{${^4}$ Department of Physics and Astronomy, University of Waterloo, Waterloo, Ontario, N2L 3G1, Canada}

\affiliation{${^6}$ Perimeter Institute for Theoretical Physics, Waterloo, Ontario, N2L 2Y5, Canada}

\date{\today}

\begin{abstract}
We report the effect of the cosmological constant and the internal energy density of a cosmic string on the deflection angle of light in the spacetime of a rotating cosmic string with internal structure. We first revisit the deflection angle by a rotating cosmic string and then provide a generalization using the geodesic equations and the Gauss-Bonnet Theorem. We show there is an agreement between the two methods when employing higher order terms of the linear mass density of the cosmic string. By modifying the integration domain for the global conical topology, we resolve the inconsistency between these two methods previously reported in the literature. We show that the deflection angle is not affected by the rotation of the cosmic string; however, the cosmological constant $\Lambda$ strongly affects the deflection angle, which generalizes the well-known result.
\end{abstract}

\maketitle

\section{Introduction}
The detection of gravitational waves from black holes and neutron star mergers, ushers in an age of multi-messenger astronomy \cite{ligob,ligon}. Another physically interesting  tool of general relativity is gravitational lensing, the slight warping of light from distant galaxies under the influence of a massive object, such as a planet, a black hole, or dark matter. Recently, weak gravitational lensing (WGL) has been used to detect dark matter filaments connecting individual massive clusters, creating a composite image of the bridge and contributing to our understanding of the large-scale structure in the Universe \cite{dmw}.  Moreover, WGL has also been used to observe the cosmological weak lensing effect on temperature fluctuations in the Cosmic Microwave Background (CMB) and to create sky maps of what could be called  ``the index of refraction of the entire visible universe" \cite{CMBw1,CMBw2,Planck,Namikawa:2013wda,Nutku:1977wp,Bartelmann:2016dvf}.

Here, we improve the weak deflection limit analysis of rotating cosmic strings with a cosmological constant \cite{Novello:1993an}. In seminal papers by Vilenkin \cite{Vilenkin:1984ib} and Gott \cite{gott} the deflection angle in the gravitational field of a static, cylindrically symmetric string has been investigated using a linear approximation to general relativity. Cosmic strings are topologically stable objects that may have been formed during the $U(1)$ symmetry breaking phase transition in the early universe when the universe had cooled down to a certain critical value \cite{kibble}. Recent studies show that the stochastic background of gravitational waves produced by a network of cosmic strings could be detected by Laser Interferometer Gravitational-Wave Observatory (LIGO) or future facilities, such as Laser Interferometer Space Antenna (LISA) \cite{  Brandenberger:1985tx,Vishniac:1986sk,Vollick:1991jf,Gogberashvili:1991np,Clement:1994sa,Ashoorioon:2014ipa,Damour:1996pv,Shlaer:2005gk,Brihaye:2008uy,mazur,aza1,  Polchinski:2007rg,Olum:2006ix,Vanchurin:2005pa,Wachter:2016hgi,Wachter:2016rwc,Blanco-Pillado:2017rnf,Blanco-Pillado:2017oxo,Cui:2017ufi,Mojahedi:2017ful,  Mandelbaum:2017jpr,Jee:2017boz}. This will provide evidence for cosmic strings fingerprints in the near future.

Alternatively, gravitational lensing may enable the measurement of cosmic strings. Towards this purpose, in this article we use the Gauss-Bonnet theorem (GBT) applied to the optical geometry, integrating the Gaussian curvature of the optical metric outwards from the light ray \cite{gibbons1,Gibbons:2008zi,Werner:2012rc}   and calculating the light deflection by cosmic strings \cite{Gibbons:1993cy}. This method has been used in many different spacetimes \cite{Jusufi:2015laa,Jusufi:2016wiz,Sakalli:2017ewb,GMWL1,GMWL2,GBWL3,Jusufi:2017gyu,Ono:2017pie,Jusufi:2017mav,Jusufi:2017vew, Jusufi:2017xnr}. It has been shown that the deflection of light by a rotating cosmic string is affected by a term which is proportional to the rotational cosmic string parameter, $a=4J$, and the linear mass density of the cosmic string, $\mu$, given by $ \delta \hat{\alpha}=3 \pi a \mu/2b$ \cite{Jusufi:2016wiz,GMWL2}. In \cite{GMWL2} it is argued that one can neglect this term since $a$ contains the linear mass density of the string per unit length, and hence this term is proportional to $\mu^2$; the agreement between the GBT, and the geodesic method breaks down in the above papers. In principle this problem should be resolvable properly since the geodesic equation method shows that the deflection angle is not affected by the rotation of the cosmic string.

Here, we aim to extend the GBT method to include the second order terms in $\mu$. We shall argue that the above-mentioned disagreement is due to the straight line approximation used in the above papers. We show that the problem can be resolved by modifying the integration domain for the global conical topology. 

Finally, we try to reveal the effect of the cosmological constant on the deflection of light in the cosmic string spacetime. Recently, many researchers have shown that there is an effect of the cosmological constant on deflection angles in the context of black holes \cite{cosmol1,cosmol2,cosmol3,cosmol4,cosmol5}. Whether the cosmological constant has an effect on the deflection angle is an open question. For this purpose, we study the effect of the cosmological constant on light deflection in relation to the internal structure of the string.

\section{Deflection angle using geodesic equations} 
The metric for a cosmic strings with an internal structure filled with matter and vacuum  energy can be written as \cite{Novello:1993an}
\begin{equation}\label{20}
\mathrm{d}s^{2}=-\mathrm{d}t^2+\mathrm{d}\rho ^{2}+\eta ^{2}\rho
^{2}\mathrm{d}\varphi ^{2}+\mathrm{d}z^{2},
\end{equation}
where $\eta=1-4\mu \zeta$, and $\zeta=\frac{\rho_0+\Lambda}{\rho_0}$. Note that $\mu$ is the linear mass density and $\rho_0$ is the internal energy density of the cosmic string. Introducing a rotation $\mathrm{d}t\to\mathrm{d}t+a\,\mathrm{d}\varphi$, with $a=4J$, and passing into spherical coordinates yields
\begin{equation}\label{21}
\mathrm{d}s^{2}=-\left(\mathrm{d}t+a\,\mathrm{d}\varphi\right)^2+\mathrm{d}r ^{2}+r^2\mathrm{d}\theta ^{2}+\eta ^{2}r^{2}\sin^2\theta\,\mathrm{d}\varphi ^{2}.
\end{equation}

Using the Euler-Lagrange equations, one can evaluate two conserved conjugate momenta in geodesic motion when both the observer and the source lie in the equatorial plane:
\begin{eqnarray}
p_{\varphi}&=&\frac{\partial \mathcal{L}}{\partial \dot{\varphi}}=-\left(\dot{t}+a \dot{\varphi}\right)a+\eta^2  r(s)^2 \dot{\varphi} =\mathcal{H}   \\
p_{t}&=&\frac{\partial \mathcal{L}}{\partial \dot{t}}=-\left(a \dot{\varphi}+\dot{t}\right)=-\mathcal{E},
\end{eqnarray}
where $\mathcal{E}$ is the energy at infinity and $\mathcal{H}$ is the angular momentum. We define the angle $\varphi$ to be measured from the point of closest approach, i.e., $u_{max}=1/b$, (note that $r=1/u(\varphi)$). Setting the affine parameter along light rays to unity, i.e., $\mathcal{E}=1$, and setting $\mathcal{H}=\eta  \,b$ leads to the following geodesics equation:

\begin{eqnarray}\notag\label{75}
\frac{1}{2 u^4}\left( \frac{\mathrm{d}u}{\mathrm{d}\varphi}\right)^2+\frac{\eta^2  }{2 u^2}-\frac{1}{2}\frac{\left(\eta^2 -a^2 u^2-\eta b u^2 a \right)^2}{u^4 \left(a+\eta b\right)^2} \\ 
-\frac{a \left(\eta^2-a^2 u^2-\eta b u^2 a \right)}{u^2 \left(a+\eta b\right)}-\frac{a^2}{2}=0.
\end{eqnarray}
Solving the last equation for $\varphi$ we obtain
\begin{equation}
\varphi=\int_0^{1/b} \frac{\eta b+a}{\sqrt{\eta^2-a^2 u^2-2a \eta b u^2-\eta^2 a^2 u^2}} \mathrm{d}u.
\end{equation}
Taking a series expansion around $\mu$ and $a$ in the integrand, one finds that the integral becomes singular at $u=0$ and $u=1/b$, given as
\begin{eqnarray}\notag
\varphi &=& \arctan\left(\frac{b u}{\sqrt{1-b^2u^2}}  \right)\Big[1+4 \mu \zeta+16 \mu^2 \zeta^2...\Big]\\
&+&\text{``other terms''}.
\end{eqnarray}

Hence the idea is to assign a value to this divergent integral at these singular points. To overcome this issue and to make the result finite, we simply assign a value to the last expression by taking the limit at the singular points and making use of the relation $\hat{\alpha}=2|\varphi_{u=1/b}-\varphi_{u=0}|-\pi$. In particular, the deflection angle is found to be
\begin{equation}
\hat{\alpha} = 4 \pi \mu \left(\frac{\rho_0+\Lambda}{\rho_0}\right)+16 \pi \mu^2 \left(\frac{\rho_0+\Lambda}{\rho_0}\right)^2+....
\end{equation}
In the special case when the cosmological constant vanishes, i.e., $\Lambda=0$, we find $\eta=1-4 \mu$, and $\hat{\alpha} = 4\pi\mu+16 \pi \mu^2+...$. For a typical ground unified value we take $\mu \simeq 10^{-6}$.

\section{Deflection angle using the Gauss-Bonnet Theorem} 
It is straightforward to show that the optical metric of the rotating cosmic string spacetime (2) is given by a Finslerian optical metric, also known as a Randers metric \cite{randers}, with a corresponding Hessian of the form (see, for example, \cite{Werner:2012rc})
\begin{equation}
g_{ij}(x,v)=\frac{1}{2}\frac{\partial ^{2}\mathcal{F}^{2}(x,v)}{\partial v^{i}\partial
v^{j}}.
\end{equation}
It is well-know that the Randers metric \cite{randers} can be written as
 $\mathcal{F}(x,v)=\sqrt{\alpha_{ij}(x)v^{i}v^{j}}+\beta_{i}(x)v^{i}$, with $\alpha_{ij}$ and $\beta_{i}$ satisfying $\alpha^{ij}\beta_{i}\beta_{j}<1$. Once we find the Randers metric, we need to apply the so-called Naz{\i }m's construction to find the Riemannian manifold $(\mathcal{M},\bar{g})$, which osculates relative to the Randers-cosmic string metric $(\mathcal{M},F)$. One can do this by simply choosing a smooth and non-zero vector field $\bar{v}$ with the Hessian $\bar{g}_{ij}(x)=g_{ij}(x,\bar{v}(x)).$  Note that the choice of vector field is not unique; for the purpose of our problem we shall take $\bar{v}^{r}=-\cos \varphi $, and $ \bar{v}^{\varphi }=\sin ^{2}\varphi/b$. 
 
We can now proceed to apply the GBT. For simplicity, we choose the origin of our coordinate system perpendicular to the cosmic string. Further, we need to chose a domain of integration, $\mathcal{D}_{R}$, a region in the equatorial
plane of the osculating Riemannian manifold bounded by the light ray $\gamma _{\bar{g}}$, at a distance $b$ from the cosmic string, and a circular segment $C_R$ of radius $R$ ( i.e., $\partial \mathcal{D}%
_{R}=\gamma _{\bar{g}}\cup C_{R}$) \cite{Werner:2012rc}. The GBT can then be stated as
\begin{equation}
\iint\limits_{\mathcal{D}_{R}}K\,\mathrm{d}S+\oint\limits_{\partial \mathcal{%
D}_{R}}\kappa \,\mathrm{d}t+\sum_{i}\theta _{i}=2\pi \chi (\mathcal{D}_{R}),
\label{32}
\end{equation}
where $K$ is the Gaussian optical curvature and $\kappa$ is the geodesic curvature defined as $\kappa =|\nabla _{\dot{\gamma}}\dot{%
\gamma}|$ \cite{Jusufi:2015laa}. In the limit as $R$ goes to infinity, the geodesics curvature reduces to  a simple form: 
\begin{equation}
\lim_{R\rightarrow \infty }\kappa (C_{R})\mathrm{d}t=\lim_{R\rightarrow
\infty }\left( \eta -\frac{a}{R}\right) =\eta \,\mathrm{d}\,\varphi .  \label{35}
\end{equation}
Clearly, when $\mu=\Lambda=0$, yielding $\eta \rightarrow 1$, we obtain the asymptotically
Euclidean case, i.e., $\kappa (C_{R})\mathrm{d}t/\mathrm{d}\varphi =1$. 
The deflection angle in the form of the GBT becomes 
\begin{equation}
\hat{\alpha}= \mathcal{I}-\frac{1%
}{\eta }\int\limits_{0}^{\pi }\int\limits_{r_\gamma}^{\infty }K\,\sqrt{\det \bar{g}}\,\mathrm{d}r\,\mathrm{d}\varphi,
\label{36}
\end{equation}
where
\begin{equation}
\mathcal{I}=\pi \left( \frac{1}{\eta} -1\right). \label{I}
\end{equation}
The determinant of the metric, neglecting higher order terms of the angular momentum parameter $a$, can be written as
\begin{equation}
\det \bar{g}=r^2 \eta^2 - \frac{3 a\sin^2 \varphi \eta^2 r^2 (\sin^4 \varphi  \eta^2 r^2 +\cos^2 \varphi b^2)}{b^3 \left(\cos^{2}\varphi+\frac{r^2 \eta^{2} \sin^{4}\varphi }{b^2}\right)^{3/2}}.
\end{equation}
Note that the components of the optical metric of the rotating cosmic string are
\begin{align}\label{39}
\bar{g}_{rr}&=1-\frac{ \sin^{6}\varphi  \eta^2 r^2 a }{b^3 \left(\cos^{2}\varphi+\frac{r^2 \eta^{2} \sin^{4}\varphi }{b^2}\right)^{3/2}}+\mathcal{O}(a^2),\\
\bar{g}_{\varphi \varphi}&=r^2 \eta^2 -\frac{a\sin^2 \varphi  r^2 \left(2 r^2 \eta^2 \sin^4 \varphi +3 b^2 \cos^2 \varphi \right)\eta^2 }{b^3 \left(\cos^{2}\varphi+\frac{r^2 \eta^{2}  \sin^{4}\varphi }{b^2}\right)^{3/2}},\\
\bar{g}_{r\varphi}&=\frac{a \cos^{3}\varphi}{b^3 \left(\cos^{2}\varphi+\frac{r^2 \eta^{2}  \sin^{4}\varphi }{b^2}\right)^{3/2}}+\mathcal{O}(a^2).
\end{align}
The Gaussian curvature is
\begin{equation}
K=\frac{1}{\sqrt{\det \bar{g}}}\left[\frac{\partial}{\partial \varphi}\left(\frac{\sqrt{\det \bar{g}}}{\bar{g}_{rr}}\,\bar{\Gamma}^{\varphi}_{rr}\right)-\frac{\partial}{\partial r}\left(\frac{\sqrt{\det \bar{g}}}{\bar{g}_{rr}}\,\bar{\Gamma}^{\varphi}_{r\varphi}\right)\right],
\end{equation}
so, using the Christoffel symbols and the metric components, we obtain
\begin{equation}
K=-\frac{12 a }{r }f(r,\varphi,\eta),\label{26}
\end{equation}
with

\begin{equation}
f(r,\varphi,\eta)= \frac{\sin^3 \varphi}{\left(\cos^{2}\varphi+\frac{r^2 \eta^{2} \sin^{4}\varphi }{b^2}\right)^{7/2}b^7} \times \end{equation}

\begin{equation}
\Big[-\frac{\sin^{11}\varphi  \eta^6 r^5}{24}+ \frac{b^2 r^3 \eta^2 \sin^9 \varphi}{8}+\frac{r^3 \eta^2  \cos^{2}\varphi b^2 (\eta^2+27) \sin^7 \varphi}{24} \notag
\end{equation}

\begin{equation}
-\frac{3 b^3 r^2 \eta^2  \cos^2 \varphi \sin^6 \varphi }{4}
+\Big(\frac{5 b^2 r^3 \eta^2  \cos^4 \varphi}{4}-\frac{\cos^2 \varphi b^4 r}{2}\Big)\sin^5 \varphi \notag \end{equation}

\begin{equation}
- \frac{3 r^2 b^3 \eta^2 \cos^4 \varphi \sin^4 \varphi }{2}+ \frac{17 r (\eta^2 -\frac{33}{17}) \cos^4 \varphi b^4 \sin^3 \varphi}{24}
\notag \end{equation}

\begin{equation}
+\frac{\cos^4 \varphi \sin^2 \varphi b^5}{2}-\frac{5 b^4 r \sin \varphi \cos^6 \varphi}{4}+b^5 \cos^6 \varphi \Big].  \notag
\end{equation}
The deflection angle \eqref{36} becomes
\begin{equation}
\hat{\alpha}= \mathcal{I}-\frac{1}{\eta}\int\limits_{0}^{\pi}\int\limits_{r_{\gamma}(\varphi)}^{\infty}\left(-\frac{12a}{r}f(r,\varphi,\eta)\right)\sqrt{\det \bar{g}}\,\mathrm{d}r\,\mathrm{d}\varphi.
\end{equation}
We then linearize Eq. \eqref{75} around $a$:
\begin{equation}
\left(\frac{\mathrm{d}u}{\mathrm{d}\varphi}\right)^2 \frac{1}{2 u^4}+\frac{\eta^2}{2 u^2}-\frac{\eta^2}{2 b^2 u^4}+\frac{a\eta^2}{u^4 b^3}=0.
\end{equation}
Differentiating this equation, we end up with 
\begin{equation}
\frac{\mathrm{d}^2}{\mathrm{d}\varphi^2}u(\varphi)+\eta^2 u(\varphi)=0.
\end{equation}
Solving this equation, one finds
\begin{equation}
u(\varphi)=C_1 \sin\left(\eta \varphi\right)+C_2 \cos\left(\eta \varphi\right).
\end{equation}
We make use of the following initial conditions: $u(\varphi=0)=0$ and $u(\varphi=\pi/2)=1/b$. It follows that
\begin{equation}
u(\varphi)=\frac{\sin\left(\eta \varphi\right)}{b}\frac{1}{\sin\left(\frac{\eta \pi}{2}    \right)}.
\end{equation}
We can approximate this equation as
\begin{equation}
u(\varphi)\simeq \frac{\sin\left(\eta \varphi\right)}{b},
\end{equation}
since $\sin\left(\frac{\eta \pi}{2}\right)\simeq 1 $. Finally, in terms of the radial coordinate $r$, we find
\begin{equation}
r(\varphi)=\frac{1}{u(\varphi)}=\frac{b}{\sin\left(\eta \varphi\right)}=\frac{b}{\sin\left[\left(1-4\mu\zeta\right) \varphi\right]}.
\end{equation}
Taking this result into the consideration, we can find no contribution from the rotating cosmic string parameter to the deflection angle:
\begin{eqnarray}
\int\limits_{0}^{\pi}\int\limits_{\frac{b}{\sin\left[\left(1-4\mu\zeta\right) \varphi\right]}}^{\infty} \frac{12  a}{r}f(r,\varphi,\eta)\sqrt{\det \bar{g}}\,\mathrm{d}r\,\mathrm{d}\varphi=0.
\end{eqnarray}
As we can see from the last equation, this simple modification of the integration domain resolves the inconsistency between GBT and the geodesic methods reported in \cite{GMWL2,Jusufi:2016wiz}. On the other hand, going back to Eq. \eqref{I}, we find
\begin{equation}
\mathcal{I}=\pi \left( \frac{1}{1-4\mu\zeta} -1\right),
\end{equation}
by expanding 
\begin{equation}
\frac{1}{1-4\mu\zeta}=1+4\mu\zeta+...,
\end{equation}
provided $4\mu\zeta <1$. Otherwise, in the case $4\mu\zeta>1$, the series diverges. Hence the total deflection angle in our case is
\begin{equation}
\hat{\alpha} = 4 \pi \mu \left(\frac{\rho_0+\Lambda}{\rho_0}\right)+16 \pi \mu^2 \left(\frac{\rho_0+\Lambda}{\rho_0}\right)^2+....
\end{equation}
We have shown that the total deflection angle is corrected and the effect of the cosmological constant has been included. 

\section{Conclusions} 

In summary, we have studied the WGL of rotating cosmic strings using the GBT and geodesics. Although cosmic strings have been known for over 40 years, they have never been observed. We have provided a detailed analysis of the influence of the cosmological constant and cosmic string parameters on the deflection angle; we have shown that, by carefully choosing the domain of integration, a generalized deflection angle containing higher order terms in the linear mass density can be obtained. When the cosmological constant $\Lambda$ vanishes, we obtain the well-known deflection angle of rotating cosmic strings.

Clearly, the deflection angle with the effect of the cosmological constant contains a clue regarding dark energy. The main message of this article is that, besides the effect of linear mass density on the deflection angle, the deflection angle is also affected by the internal structure of the string (such as the vacuum energy, as given by the cosmological constant). On the other hand, we have shown that there is no contribution from the rotation of the cosmic string, when higher order terms in $\mu$ are included. The effect is larger as a result of the cosmological constant $\Lambda$ and energy density $\rho_0$, which might play an important role at astrophysical scales. These results provide an excellent opportunity to observe cosmic strings by WGL and to determine the nature of these exotic objects. Hence, gravitational lensing can provide some clues towards the direct detection of the nature of the cosmological constant and cosmic strings.

\subsection*{Acknowledgements} 

A\"{O}~acknowledges financial support provided under the Chilean FONDECYT Grant No. 3170035. A\"{O} is grateful to the Waterloo University, Department of Physics and Astronomy, and to the Perimeter Institute for Theoretical Physics for hosting him as a research visitor.

\end{document}